\documentclass[aps,prb,reprint,a4paper,amsmath,amssymb,graphicx,superscriptaddress,longbibliography,floatfix]{revtex4-1}
\usepackage[utf8]{inputenc}
\usepackage[english]{babel}
\usepackage{graphicx}
\usepackage[separate-uncertainty=true]{siunitx}
\usepackage{xcolor}
\usepackage[
	pdffitwindow=true,
	colorlinks=true,
	frenchlinks=false,
        linkcolor=violet,
	anchorcolor=violet,
        citecolor=violet,
        filecolor=violet,
        urlcolor=violet,
        bookmarks=true,
        bookmarksopen=true,
	bookmarksnumbered=true, 
        bookmarksopenlevel=1,
        plainpages=false,
	pdfpagelayout=TwoPageLeft,
        pdfpagelabels=true,
	breaklinks
]{hyperref}

\newcommand{\BZP}{B_\mathrm{1,0}}
\newcommand{\geff}{g_\mathrm{eff}}
\newcommand{\kext}{\kappa_\mathrm{ext}}
\newcommand{\gammas}{\gamma_\mathrm{s}}
\newcommand{\muB}{\mu_\mathrm{B}}
\newcommand{\Ryz}{R_{yz}}

\begin{document}

\title{Quantitative modeling of superconducting planar resonators with improved field homogeneity for electron spin resonance}

\author{Stefan Weichselbaumer}
\email[]{stefan.weichselbaumer@wmi.badw.de}
\affiliation{Walther-Meißner-Institut, Bayerische Akademie der Wissenschaften, 85748 Garching, Germany}
\affiliation{Physik-Department, Technische Universität München, 85748 Garching, Germany}

\author{Petio Natzkin}
\affiliation{Walther-Meißner-Institut, Bayerische Akademie der Wissenschaften, 85748 Garching, Germany}
\affiliation{Physik-Department, Technische Universität München, 85748 Garching, Germany}

\author{Christoph W.\ Zollitsch}
\altaffiliation[Present address: ]{London Centre for Nanotechnology, UCL, London WC1H 0AH, United Kingdom}
\affiliation{Walther-Meißner-Institut, Bayerische Akademie der Wissenschaften, 85748 Garching, Germany}
\affiliation{Physik-Department, Technische Universität München, 85748 Garching, Germany}

\author{Mathias Weiler}
\affiliation{Walther-Meißner-Institut, Bayerische Akademie der Wissenschaften, 85748 Garching, Germany}
\affiliation{Physik-Department, Technische Universität München, 85748 Garching, Germany}

\author{Rudolf Gross}
\email[]{rudolf.gross@wmi.badw.de}
\affiliation{Walther-Meißner-Institut, Bayerische Akademie der Wissenschaften, 85748 Garching, Germany}
\affiliation{Physik-Department, Technische Universität München, 85748 Garching, Germany}
\affiliation{Nanosystems Initiative Munich, 80799 München, Germany}

\author{Hans Huebl}
\email[]{hans.huebl@wmi.badw.de}
\affiliation{Walther-Meißner-Institut, Bayerische Akademie der Wissenschaften, 85748 Garching, Germany}
\affiliation{Physik-Department, Technische Universität München, 85748 Garching, Germany}
\affiliation{Nanosystems Initiative Munich, 80799 München, Germany}

\date{\today}

\begin{abstract}
We present three designs for planar superconducting microwave resonators for
electron spin resonance (ESR) experiments. We implement finite element
simulations to calculate the resonance frequency and quality factors as well as
the three-dimensional microwave magnetic field distribution of the resonators.
One particular resonator design offers an increased homogeneity of the microwave
magnetic field while the other two show a better confinement of the mode volume.
We extend our model simulations to calculate the collective coupling rate
between a spin ensemble and a microwave resonator in the presence of an
inhomogeneous magnetic resonator field. Continuous-wave ESR experiments of
phosphorus donors in $^\mathrm{nat}$Si demonstrate the feasibility of our
resonators for magnetic resonance experiments. We extract the collective
coupling rate and find a good agreement with our simulation results,
corroborating our model approach. Finally, we discuss specific application cases
for the different resonator designs.
\end{abstract}

\maketitle

\section{Introduction}
Microwave resonators are a key part of any ESR experiment. They enhance the
microwave magnetic field at the sample location and offer equally enhanced
sensitivity for inductive detection of magnetization
dynamics\cite{Poole1996,Schweiger2001}. While conventional ESR resonators based
on three-dimensional (3D) microwave cavities provide a microwave magnetic field
with high homogeneity over a large volume, they suffer from small filling
factors and, in turn, a low sensitivity for small samples. 
Planar microresonators allow to reduce the mode volume, which, depending on the
sample size and geometry, can lead to an increased filling factor and therefore
an enhanced sensitivity compared to 3D
cavities\cite{Narkowicz2005,Narkowicz2008,Torrezan2009}.
In addition, planar resonators operated at low temperatures allow one to use
superconducting materials, offering small losses and extraordinarily high
quality factors. Making use of these advantages led to a plethora of planar
resonator geometries\cite{Malissa2013,Benningshof2013,Sigillito2014}, applicable
in several different fields of expertise. In addition, superconducting
resonators have also become key components in the field of circuit quantum
electrodynamics (cQED)\cite{Devoret2013} and led to a subsequent introduction of
cQED concepts in the field of magnetic
resonance\cite{Wesenberg2009,Wu2010,Abe2011,Sandner2012}. The quest for
ultra-sensitive ESR at low temperatures has led to a range of experiments.
Well-known examples are the use of parametric amplification based on
superconducting quantum circuits\cite{Bienfait2016,Probst2017} or the use of
quantum states as a resource to increase the signal-to-noise
ratio\cite{Bienfait2017}. Another direction is to increase the coupling rate
between a spin ensemble and the microwave resonator to enhance the read-out
sensitivity of the measurements. Here, the so-called strong coupling regime has
been achieved for several types of spin systems in combination with
superconducting
resonators\cite{Kubo2010,Schuster2010,Amsuss2011,Probst2013,Zollitsch2015}.

Despite inspring progress in ultra-sensitive ESR, so far a
quantitative analysis of planar resonator designs and their suitability for achieving
strong coupling and enabling straight forward coherent control of spin systems
is still missing. Here, we employ finite element simulations of superconducting
planar microwave resonators for calculating the spatial distribution of the
microwave magnetic field of three different resonator geometries. This
information is crucial to judge the performance of the resonator for the
specific application. We demonstrate that these simulations can not only be
used to predict the resonance frequencies and quality factors but also allow for
a quantitative comparison of the magnetic field homogeneity. Additionally, the
simulated magnetic field distribution enables us to calculate the expected
collective coupling rate between the spin ensemble and the microwave resonator.
Finally, we compare our model predictions to actual continuous-wave ESR data and
find a good agreement between theory and experiment, including the modeling of
power-dependent saturation effects. Here, we show that the modified
power saturation is confirming the simulations of the microwave
magnetic field distribution.
In addition, this work compares the various resonator designs with different
levels of microwave magnetic field homogeneity. This is relevant in the context
of pulsed ESR experiments, as homogeneous microwave magnetic excitation fields
are a key requirement for the coherent control of spin ensembles using
rectangular pulse excitation schemes. 

The experimental data presented in this work is recorded at a
temperature of \SI{1.5}{K}, i.e.\ the resonator is not in its quantum ground
state. Chiorescu~\emph{et~al.} demonstrated by numerical simulations that a
transition to the classical spin-resonance mechanism occurs when the number of
photons in the resonator, $n_\mathrm{ph}$, is large compared to the number of
spins, $N$\cite{Chiorescu2010}. However, as we will show later, this is not the
case for the measurements presented in this manuscript. Our data therefore
allows to compare the computed effective coupling between the microwave
resonator and the spin ensemble also at Millikelvin temperatures, as the thermal
spin polarization can be taken into account\cite{Zollitsch2015}.

The paper is organized as follows. First, we introduce three different planar
resonator designs  and present important design considerations. We then
introduce the concept of spin-photon coupling in the context of electron spin
resonance and show how the collective coupling rate can be computed in the
presence of an inhomogeneous microwave magnetic field. Subsequently, we present
our simulation approach. We quantitatively analyze the field homogeneity of two
of the resonator designs and show that one particular design offers an improved
field homogeneity. In the following experimental section, we first confirm the
feasibility of our simulation approach. The second part of the experimental
section is dedicated to continuous-wave ESR experiments on an ensemble of
phosphorus donors in silicon. We extract the collective coupling rate and find a
good agreement between the theoretical model and the experimental data. Finally,
we also model the power-dependent saturation of the collective coupling rate.

\section{Theoretical considerations}
\subsection{Microwave resonator designs}
\begin{figure}
	\center
	\includegraphics[width=0.90\linewidth]{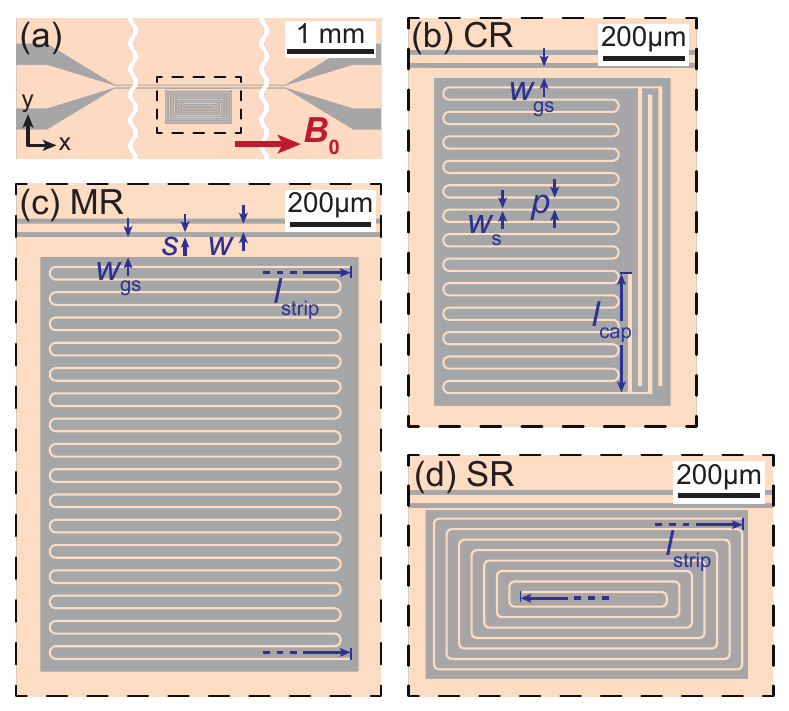}
    \caption{\label{SimulationSetup}Schematic illustration of the resonator
        designs studied in this work. For details and parameters, see text.
        (a)~Generic device layout. A central feedline with two connection pads
        excites the resonators which are placed in the proximity of the feedline
        (dashed box). 
        (b)~Capacitively-shunted meander resonator (CR) consisting of a
        interdigitated finger capacitor shunting a meandering inductor.
        (c)~The meander resonator (MR) is a half-wavelength resonator consisting
        of a long meander-shaped strip.
        (d)~For the spiral resonator (SR), the meander-shaped strip is arranged
        in a two-dimensional coil-like structure, providing an increased field
        homogeneity.		
		}
\end{figure}

In the following, we present the sample layout and the resonator designs
presented in this article.  Fig.~\ref{SimulationSetup}\,(a) displays the generic
design of a chip featuring a central feedline, designed in coplanar waveguide
geometry\cite{Wen1969}, with two connection pads at the edges of the substrate.
More details and characteristic parameters are given in
Section~\ref{sec:simulation_setup}.

The planar microwave resonator structure is placed in proximity to a microwave
feedline (dashed box). Figure~\ref{SimulationSetup}\,\mbox{(b)--(d)} shows the
three discussed resonator designs, which are (b) a capacitively-shunted meander
resonator (CR), (c) a meander  resonator (MR), and (d) a spiral resonator (SR).
For the MR, the capacitance is provided by the intra-line capacitance with a
higher inductance to capacitance ratio than the CR. The SR design also relies on
the intra-line capacitance, but provides a more homogeneous magnetic field
distribution than the MR design. 

The CR is a so-called lumped element microwave
resonator\cite{Lindstrom2009,Khalil2011,Geerlings2012} where the capacitance and
the inductance are most obviously visible in form of the interdigital capacitor
and the meandering inductor. Nevertheless, a numeric analysis of the resonance
frequency as a function of the interdigital capacitance finger length (see
Fig.~\ref{SimulationComparison}\,(a)) suggests that there is a finite
capacitance contained in the meandering structure for the design displayed. In
consequence, the meandering resonator is the logical next step, where the
interdigital capacitor is completely missing. The explicit missing capacitance
is compensated by an enhanced total length of the resonator structure. An
alternative viewpoint is to think of the meander resonator as a waveguide
structure, where the length of the conductor supports a standing wave
pattern of the microwave\,\cite{Wallace1991a}. Here, the length of the
meander-shaped strip of the MR and SR corresponds to half of the wavelength of
the resonance frequency and thus supports this picture.

The meander shape of the CR and MR results in a counter-flow of the
high-frequency currents in neighboring meandering strips. This ennables a
localization of the electromagnetic field close to the surface. In addition,
this also leads to a significant inhomogeneity of the microwave magnetic field
$B_1$ in close proximity of the structure. Moreover, a fast decay of the
microwave excitation $B_1$ field in the $z$-direction, i.e.\ out of the plane of
the microwave resonator (see Fig.~\ref{FieldDistribution}\,(a) and (c)) is
obtained. The characteristic decay length of the $B_1$ field in $z$-direction is
related to the distance between adjacent wires, which is \SI{20}{\mu m} in our
design.  Although this can be beneficial for the measurement of ultra-thin spin
samples, for most pulsed ESR experiments, using rectangular pulse excitation
schemse, the large field inhomogeneity is
undesired and elaborate techniques have been proposed to compensate the $B_1$
inhomogeneity\cite{Kupce1995,Tannus1997,Garwood2001,Skinner2011,Spindler2012}. A
suitable resonator design leading to a significant reduction of the magnetic
field inhomogeneity comes in the form of a spiral resonator geometry as
displayed in Fig.~\ref{SimulationSetup}\,(d). Here, neighboring lines have a
parallel flow of current resulting in a much better homogeneity of the $B_1$
field as shown in Fig.~\ref{FieldDistribution}\,(b) and (d). Note that the field
extends now significantly further into the $z$-direction and the characteristic
decay length is in the same order of magnitude as the lateral dimensions of the
whole resonator.

The design resonance frequency $f_r$ of the superconducting microwave resonators
discussed here is set to a value around \SI{5}{GHz}. This optimizes the surface
impedance/losses of the structure (which increase with increasing
frequency\cite{Turneaure1991}), while keeping a reasonably high-frequency.
Additionally, our experimental setup is designed to operate in the frequency
band of 4--\SI{8}{GHz}. In general, the resonance frequency of a LC-oscillator
is given by $f_r = 1/(2\pi\sqrt{LC})$ with an effective inductance $L$ and
capacitance $C$.  Changing the length of the capacitor finger $l_\mathrm{cap}$
for the CR or the total inductor length $l_\mathrm{strip}$ for the MR and SR
allows to tune the resonance frequency to the desired value.

A further key parameter of a microwave resonator is the quality factor $Q$,
which is given by $Q = f_r/(2\kappa/2\pi)$, where $\kappa/2\pi$ is the loss rate
of the resonator (measured as the half-width at half maximum of the resonance
line).  One can distinguish between internal and external losses with
corresponding quality factors given by
\begin{equation}
	\frac{1}{Q} = \frac{1}{Q_\mathrm{ext}} + \frac{1}{Q_\mathrm{int}}.
\end{equation}
The quality factors are linked to the external and internal loss rate according
to $Q_\mathrm{ext} = f_r/(2\kext/2\pi)$ and $Q_\mathrm{int} =
f_r/(2\kappa_\mathrm{int}/2\pi)$. Internal losses, including radiation,
resistive and dielectric losses\cite{Goppl2008}, are typically very small in
superconducting resonators and internal quality factors above $10^7$ have been
reported\cite{Megrant2012}. The external loss rate $\kext$ describes the
coupling to the ``environment'', which is here the feedline. Technically, this
coupling can be either of mainly capacitive or inductive nature, depending on
wether the contact point of the resonator is close to an anti-node of the
electric or magnetic field. It can be controlled by the separation between the
resonator and the feedline. Table~\ref{tab:resonators} summarizes the
geometric parameters as well as the resonance frequency and quality factors of
the resonators presented in this work.

\subsection{Spin-photon coupling in electron spin resonance}
We now turn to the effective coupling between the spin ensemble and the
microwave magnetic field mode provided by the resonator. The vacuum coupling
strength $g_0$ between a single spin and the electromagnetic modes of a
microwave resonator is given by\cite{Wesenberg2009} 
\begin{equation}
    g_0 = g_\mathrm{s}\muB\BZP/2\hbar,
    \label{eq:single_spin_coupling}
\end{equation}
where $g_\mathrm{s}$ is the electron g-factor of the spin ensemble, $\muB$ is
the Bohr magneton and $\BZP$ describes the zero-point or vacuum fluctuations of
the magnetic field inside the resonator.  Assuming a homogeneous microwave field
distribution, $\BZP$ can be expressed as\cite{Schoelkopf2008} $\BZP =
\sqrt{\mu_0\hbar\omega_r/(2V_\mathrm{m})}$, where $\mu_0$ is the vacuum
permeability, $\hbar$ is the reduced Planck constant, and $V_\mathrm{m}$ is the
mode volume of the resonator.  Applying this to a typical coplanar
waveguide resonator with a signal line width in the order of \SI{10}{\mu m} and a
resonance frequency in the order of \SI{3}{GHz}, $g_0$ can be roughly estimated
to \SI{10}{Hz}\cite{Blais2004,Schoelkopf2008,Grezes2015a}. Nevertheless and as
discussed below, lumped element resonators typically have a complex spatially
dependent microwave field distribution which has to be taken into account.
Increasing the coupling increases the sensitivity of ESR measurements and
ultimately allows for a high-cooperativity  or even strong
coupling\cite{Tosi2014}. When considering not only a single spin, but a spin
ensemble with $N$ non-interacting spins, the interaction between the whole spin
ensemble and the resonator can be improved by making use of collective coupling
effects\cite{Dicke1954}, which predict an enhancement by a factor of $\sqrt{N}$
leading to an effective coupling strength $\geff = g_0\sqrt{N}$. The collective
coupling strength for a homogeneous $\BZP$ distribution can then be written
as\cite{Huebl2013}
\begin{equation}
	g_\mathrm{eff,hom} = \frac{g_s\muB}{2\hbar}\sqrt{\frac{1}{2}\mu_0\hbar\omega_r\rho_\mathrm{eff}\nu}.
	\label{eq:g_eff_hom}
\end{equation}
Here, we have substituted the number of spins $N$ by $N = \rho_\mathrm{eff}V =
\rho P(T) V$, where $\rho$ is the spin density, $P(T)$ is the thermal
polarization of the spin ensemble's transition and $V$ is the volume of the spin
sample\cite{Zollitsch2015}. The filling factor $\nu = V/V_\mathrm{m}$ defines
the ratio of the Si:P crystal volume to the mode volume of the resonator. It is a
crucial parameter in ESR experiments, as the detected ESR signal is directly
proportional to the filling factor\cite{Poole1996}.

We note again that Eq.~\eqref{eq:g_eff_hom} assumes a homogeneous
distribution of the $B_1$ field over the Si:P crystal. Moreover, the
equation does not consider the orientation of the static magnetic field $B_0$
relative to the $B_1$ field required for exciting ESR transitions, i.e.\
$B_0\perp B_1$\cite{Poole1996}. 

As shown in Appendix~\ref{sec:g_eff_derivation}, the inhomogeneity can be
accounted for by integrating the microwave magnetic field over the Si:P crystal
and the cavity volume, respectively, resulting in
\begin{equation}
	\nu = \frac{\int_\mathrm{Sample} |B_1(\vec{r})|^2\,\mathrm{dV}}{\int_\mathrm{Cavity}|B_1(\vec{r})|^2\,\mathrm{dV}}.
    \label{eq:filling_factor_hom}
\end{equation}
In our finite element simulations, we compute the $B_1$ spatial distribution and
export it in discrete volume
elements $\Delta V$. Thus rewriting Eq.~\eqref{eq:filling_factor_hom} in the form
of a Riemann sum allows us to numerically compute the filling factor and hence
the effective coupling. In detail, we derive an expression, which accounts for
both, the inhomogeneity of the microwave magnetic field and the excitation
condition (for ESR excitation, $\vec{B}_1$ and $\vec{B}_0$ are chosen
perpendicular):
\begin{equation}
	g_\mathrm{eff,inhom} = \frac{g_s\muB}{2\hbar}\sqrt{\frac{1}{2}\mu_0\hbar\omega_r\rho_\mathrm{eff}\Ryz}.
	\label{eq:g_eff_inhom}
\end{equation}
This expression is equivalent to Eq.~\eqref{eq:g_eff_hom}, replacing the
filling factor $\nu$ by the term
\begin{equation}
	\Ryz = \frac{\sum_{V} |B^{yz}_{1,\mathrm{sim}}(\vec{r})|^2 }{ \sum_{V_\mathrm{m}} |B^{xyz}_{1,\mathrm{sim}}(\vec{r})|^2 }.
	\label{eq:filling_factor}
\end{equation}
Here, the sum in the numerator accounts for all numerically computed microwave
magnetic field amplitudes in the Si:P crystal volume $V$, which fulfill the
excitation condition required for exciting an ESR transition. For our chosen
geometry this is the field amplitude in the $yz$-plane $\left|B_1^{yz}\right| =
\sqrt{(B_1^y)^2 + (B_1^z)^2}$. The denominator can be understood as a
normalization factor and thus accounts for the total magnetic field amplitude
$B_1^{xyz}$ in the mode volume $V_\mathrm{m}$. Both sums take the field
amplitudes $B_1^{yz}$ and $B_1^{xyz}$ at all available volume elements $\Delta
V$ computed by the FEM simulations (see Section~\ref{sec:simulation_setup}) into
account. Here, we only consider the half-space above the substrate to be filled
with the spin ensemble, thus $\Ryz$ is naturally limited to $\nu = 0.5$. For
realistic resonator structures, this value is further reduced when components of
the $B_1$ field are aligned parallel to the static magnetic field and thus do
not contribute to an ESR excitation.

Eq.~\eqref{eq:g_eff_inhom}~and~\eqref{eq:filling_factor} allow us to
calculate and theoretically predict the achievable collective coupling strength
of a spin ensemble coupled to an arbitrary microwave resonator geometry, as long
as the magnetic field distribution is known. Note that
Eq.~\eqref{eq:g_eff_inhom} in combination with \eqref{eq:filling_factor} also
includes effects originating from thermal polarization, as $\rho_\mathrm{eff} =
\rho P\left(T\right)$. In the following, we will calculate $\geff$ using the
magnetic field distribution obtained for our resonator geometries using FEM,
present experimental data for those resonators and hereby corroborate our
theoretic model.

\section{Finite element simulations}
\subsection{Simulation setup}
\label{sec:simulation_setup}
For our FEM simulations we use the commercial microwave simulation software CST
Microwave Studio 2016\cite{CSTMicrowaveStudio2016}. Technically, our modeling
takes the entire chip into account. We start with the definition of the
substrate material (here: silicon) with the dimensions of
$\SI{6}{mm}\times\SI{10}{mm}\times\SI{0.525}{mm}$. On top of the substrate, we
model the superconducting film by a \SI{150}{nm} thick perfect electrical
conductor for simplicity. Note that we do not take the kinetic
inductance\cite{Mattis1958,Turneaure1991} or the finite penetration depth into
account.
Figure~\ref{SimulationSetup}\,(a) displays the generic design of the feedline.
The width of the signal line is $w = \SI{20}{\mu m}$ and the distance between
signal line and ground plane is $s = \SI{12}{\mu m}$ corresponding to an
impedance of \SI{50}{\ohm}\cite{Niemczyk2009}. The wire thickness of the
resonator itself is $w_s = \SI{5}{\mu m}$ with a spacing of $p = \SI{20}{\mu
m}$. For the SR, the spacing is $p_x = \SI{30}{\mu m}$ in $x$-direction and $p_y
= \SI{20}{\mu m}$ in $y$-direction.

In our experiments, a spin ensemble hosted in a silicon crystal interacts with
the microwave magnetic field of the resonator. To take the finite dielectric
constant of silicon into account and to model the properties of the microwave
resonator accurately, we position a box-shaped silicon body with dimensions of
$\SI{3.4}{mm}\times\SI{3.4}{mm}\times\SI{0.42}{mm}$ on top of the resonator. In
our simulations, we apply the microwave signal to the structure via one of two
waveguide ports that are defined at both ends of the microwave feedline. The
power applied to the feedline in our simulations is $P = \SI{0.5}{W}$. In our
simulations, we consider only the linear response regime, i.e.\ any non-linear
response is not accounted for. The complete model is fully parameterized,
allowing us to efficiently explore the influence of a wide range of parameters
on the resonator parameters. 

The three-dimensional model is divided in a tetrahedral mesh cells
with a minimum  edge length of \SI{0.15}{\mu m}. During the simulation, the mesh
is adapted automatically to increase the quality of the
mesh\cite{CSTMicrowaveStudio2016}. Decreasing the mesh size further does not
lead to a change of the obtained results, therefore we conclude that our
simulations converge. The microwave magnetic field distribution is exported in a
discretized lattice with a pixel size of $1\times 1\times\SI{1}{\mu m^3}$ for the
MR/CR and $1.5\times 1.5\times \SI{1.5}{\mu m^3}$ for the SR respectively.

\subsection{Magnetic field amplitude rescaling}
In this section, we explain how we adjust the experimental and simulated
microwave magnetic field amplitude. We match the field amplitude obtained by the
simulations $B_1^\mathrm{sim}$ to the experimental conditions by rescaling
$B_1^\mathrm{sim}$ to obtain the same photon number in simulation and
experiment. The average photon number in a resonator is given by
\begin{equation}
    n_\mathrm{ph} = \frac{\kext}{\kappa^2}\frac{P_\mathrm{S}}{\hbar\omega_r},
	\label{eq:photon_number}
\end{equation}
where $P$ is the applied microwave power. We calculate a rescaling factor
$n_\mathrm{exp}/n_\mathrm{sim}$ with the average number of photons in the
resonator for the experiment, $n_\mathrm{exp}$, and the simulation,
$n_\mathrm{sim}$. For the calculations presented below, we rescale the microwave
magnetic field amplitude according to
\begin{equation}
	B_1 = B_1^\mathrm{sim}\cdot\sqrt{n_\mathrm{exp}/n_\mathrm{sim}}.
    \label{eq:rescaling_factor}
\end{equation}

\subsection{\texorpdfstring{$B_1$}{B1} magnetic field homogeneity}
\label{sec:homogeneity}
In this section, we present finite element modeling of the microwave magnetic
field distribution of the different resonator geometries and analyze the field
homogeneity with respect to a finite Si:P crystal size. For the comparison of the
field amplitude and homogeneity between the MR and SR, we rescale the magnetic
field amplitude for both resonators to an average photon number of $10^{12}$
photons, which corresponds to an input power of $\approx \SI{-19}{dBm}$ for the
SR (\SI{-14.5}{dBm} for the MR). This allows to quantitatively compare
both designs, independent of their respective quality factors.

\begin{figure}[htp]
	\center
	\includegraphics[width=\linewidth]{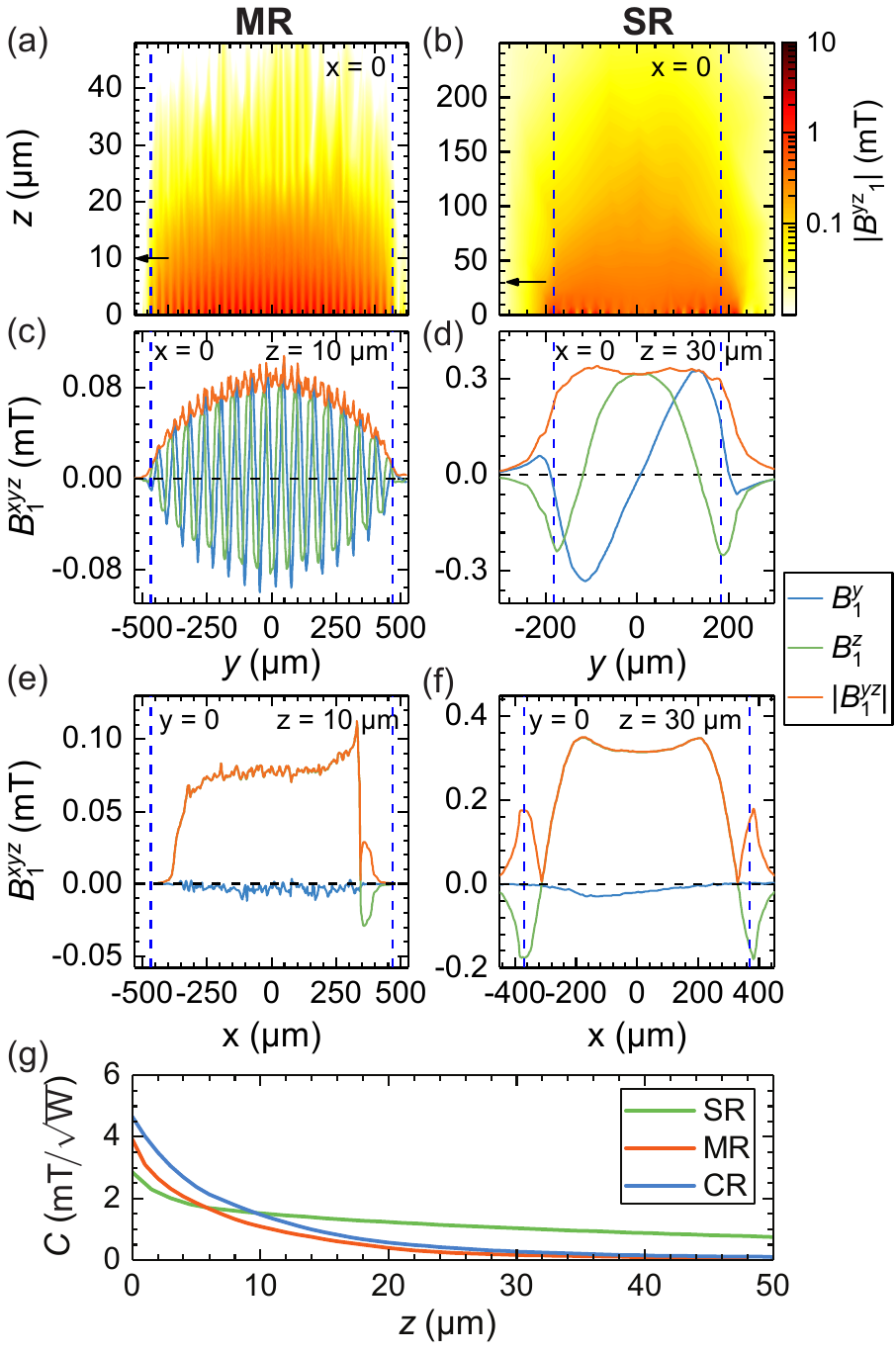}
    \caption{\label{FieldDistribution}Simulated magnetic field distribution
        $|B_1^\mathrm{yz}|$ for the (a) MR and (b) SR in the $yz$-plane.
        Anti-parallel current flow in adjacent wires results in a highly
        inhomogeneous field for the MR, while parallel current flow leads to a
        more homogeneous field for the SR. Magnetic field components $B_1^y$,
        $B_1^z$ as well as the magnitude $|B_1^{yz}|$ at fixed distance (arrows
        in panel~(a) and (b)) are plotted for the (c)~MR and (d)~SR along the
        $y$-axis and for the (e)~MR and (f)~SR along the $x$-axis.	
    (g) Conversion factor $C$ of the different resonators as a function of
    distance $z$ above the resonator for two different sample geometries (see
    text). MR and CR show larger maximum values but decrease faster with
    increasing distance compared to the SR.
	}
\end{figure}

To visualize the field homogeneity, we show the absolute magnetic field
$|B_1^\mathrm{yz}|$ in the $yz$-plane ($x = 0$) for the MR and SR in
Fig.~\ref{FieldDistribution}\,(a) and (b), respectively. Vertical dashed lines
mark the lateral extent of the resonator.  The origin in the
$xy$-plane is chosen to be the center of the resonator.

In the MR, the microwave current in adjacent wires flows anti-parallel,
resulting in opposing microwave magnetic fields. This is reflected in the
homogeneity plot in panel~(a) and is also the reason of the fast decay of the
magnetic field in the far field. In contrast, the coil-like arrangement of the
inductor wire in the SR leads to a parallel current-flow in the two halves of
the resonator and therefore to a larger homogeneity. Furthermore, the magnetic
field generated by neighboring strips does not cancel in the far field and
decays slower than for the MR (note the different scaling of the ordinate). 
To further highlight the difference between the two designs, we show cuts at a
fixed distance above the resonator (arrows in panel~(a) and (b)) for
$x = 0$ and plot the components $B_1^y$ and $B_1^z$ as well as the magnitude
$|B_1^{yz}|$ for the MR and SR along the $y$-axis in
Fig.~\ref{FieldDistribution}\,(c) and (d), respectively. The oscillatory
behavior of the magnetic field can be clearly seen for the MR, while homogeneous
excitation is obtained for the SR. In
Fig.~\ref{FieldDistribution}\,(e) and (f), we plot the field components at a
fixed distance $z$ above the resonator along the $x$-axis. Along this axis, the
field homogeneity of the MR is significantly improved.

The MR generates a maximum field of $|B_1^{yz}| = \SI{0.1}{mT}$ at a distance of
$z = \SI{10}{\mu m}$, compared to \SI{0.35}{mT} at $z = \SI{30}{\mu m}$ for the
SR. For a more detailed analysis, we evaluate the conversion factor $C$, which
is defined as\cite{Poole1996}
\begin{equation}
    C = \frac{|B_\mathrm{1,mean}^{yz}(z)|}{\sqrt{QP_\mathrm{S}}},
\end{equation}
where $Q$ is the resonator quality factor and $P_\mathrm{S}$ is the applied
microwave power. 
We assume a sample with $xy$-dimensions much larger than the
lateral extent of the resonator. $B_\mathrm{1,mean}^{yz}(z)$ is the mean
microwave magnetic field in a slice with \SI{1}{\mu m} thickness
(\SI{1.5}{\mu m} for the SR) with dimensions corresponding to the exported field
distribution.
The conversion factors for the three resonators are displayed
in Fig.~\ref{FieldDistribution}~(e) as a function of the distance $z$ above the
resonator.  
Directly above the resonator, the MR and CR show the highest
conversion factors with values up to \SI{4.6}{\mathrm{mT}/\sqrt{\mathrm{W}}}.
This value is more than one order of magnitude larger than that of commercially
available resonators.  When the distance to the resonator increases, the
conversion factor of the MR and CR decreases faster than for the SR. This is due
to the large $B_1$ inhomogeneity and the fast decay along the $z$-direction of
the MR and CR. Please note that due to the different quality factors
    of the resonators the conversion factor do not allow a direct comparison of
the obtained maximum $B_1$ amplitude.

The simulated three-dimensional field distribution can also be used to estimate
the mode volume $V_\mathrm{m}$, to the region where the field amplitude decays
to \SI{1}{\%} of its maximum value. As can be seen in
Table~\ref{tab:resonators}, the mode volume of the SR is increased compared to
the other designs, in particular in regard to the smaller lateral dimensions.

\section{Experimental Details}
\subsection{Sample fabrication and measurement setup}
\label{sec:experimental}
To fabricate the sample, we first sputter-deposit a \SI{150}{nm} thin layer of
Nb with a \SI{10}{nm} Al capping layer on a high resistivity ($>\SI{3}{k\ohm}$)
$^\mathrm{nat}$Si substrate with a thickness of \SI{525}{\mu m} (we do not take
the Al capping layer into account in our simulations). The Al capping
layer is introduced to prevent oxidation of the Nb layer. The resonators are
patterned using a standard electron beam lithography process and subsequently
etched using chemical wet etching for the Al as well as reactive ion etching for
the Nb layer. The sample chip is then placed into a copper sample holder and
connected to two SMA end launch connectors. The sample holder is mounted into a
Helium gas-flow cryostat operated at $T\approx\SI{1.5}{K}$ for all of our
experiments. 

\begin{figure}
	\center
	\includegraphics[width=\linewidth]{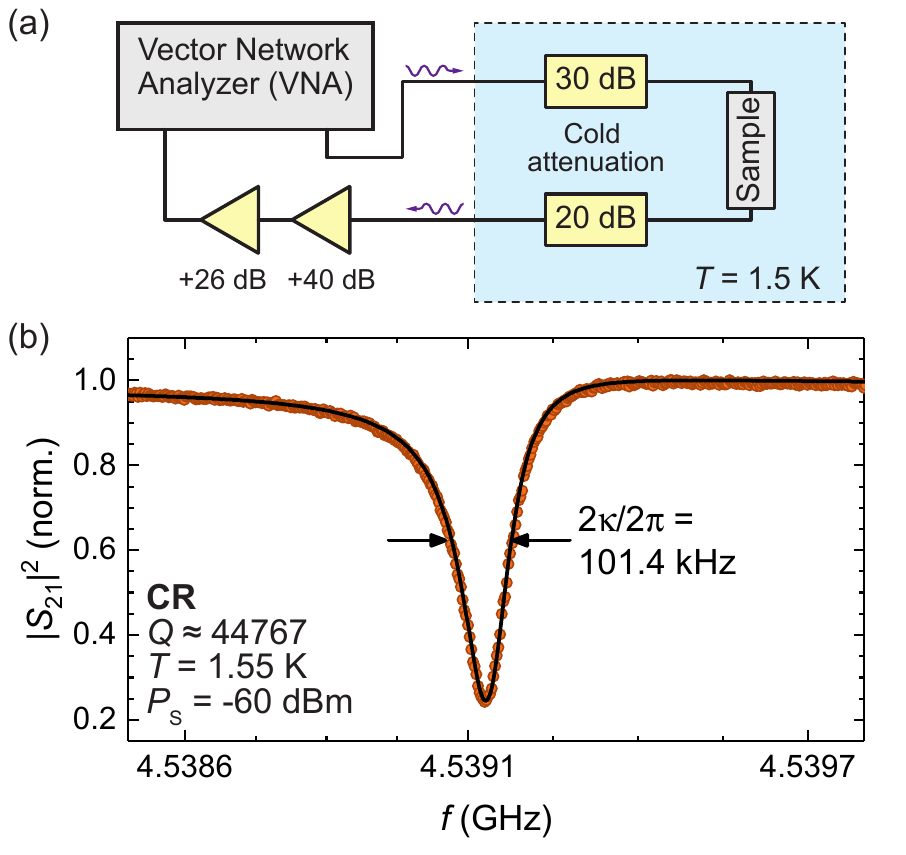}
    \caption{\label{ExperimentalSetup}(a) Schematic of the experimental setup.
        The sample holder is mounted inside a cryostat with a temperature
        $T\approx\SI{1.5}{K}$ Transmission through the sample is measured by a
        vector network analyzer (VNA). The signal is attenuated inside the
        cryostat to suppress room temperature thermal microwave photons and is
        subsequently amplified at room temperature before detection.
    (b) Exemplary transmission $|S_{21}|^2$ as a function of frequency (data
    points), recorded at zero magnetic field without a mounted Si:P crystal. A
    fitting routine (solid line) allows to extract the resonance frequency,
    linewidth and external coupling rate from the transmission dip.	
	}
\end{figure}

Figure~\ref{ExperimentalSetup}\,(a) schematically depicts the microwave
circuitry of our experiments. For the ESR experiments we apply a external
magnetic field parallel to the Nb film plane provided by a superconducting
solenoid. We measure the complex microwave transmission amplitude $S_{21}$ by
connecting the sample to the two ports of a vector network analyzer (VNA).
Hereby, we can determine the uncalibrated power transmission $|S_{21}|^2$ of the
sample. The signal is attenuated by \SI{30}{dB} (\SI{20}{dB}) in the input
(output) line inside the cryostat to avoid saturation of the ESR transitions by
room temperature thermal microwave photons. The signal is amplified by two
room-temperature low-noise amplifiers before detection. 

Figure~\ref{ExperimentalSetup}\,(b) shows an exemplary measurement, where the
transmission $|S_{21}|^2$ (points) is plotted against the frequency. The
measurement was recorded at zero magnetic field and no sample was mounted on the
chip. The attenuators were not present.  For the normalization, we set the
off-resonant transmission to one. When the excitation frequency is in resonance
with the microwave resonator, the transmission drops to about 0.15. In order to
extract the resonance frequency, the linewidth, as well as the coupling rate of
the microwave resonator to the microwave feedline from the measured complex
$S_{21}$ data, we use a robust circle fit (solid line) described by
Probst~\emph{et~al}.\cite{Probst2015}.

We summarized the relevant parameters for the resonators used in this work in
Table~\ref{tab:resonators}. 
The resonance frequency as well as the quality factors are extracted from
transmission measurements with a mounted Si:P crystal, as described above. The
extraction of the collective coupling rate, its theoretical calculation and the
estimation of the Si:P crystal-resonator gap $d_\mathrm{gap}$ is described in
Section~\ref{sec:coupling_analysis}.
\begin{table*}[tb]
	\begin{ruledtabular}
	\begin{tabular}{ l c c c c c c c c c}
		\textrm{Resonator} & Dimensions & $V_\mathrm{m}$ & $f_r$& $Q$ & $Q_\mathrm{ext}$ & $Q_\mathrm{int}$ & $g_\mathrm{eff,exp.}$ & $g_\mathrm{eff,theo.}$ & $d_\mathrm{gap}$ \\		
		 & ($\mathrm{\mu m}^2$) & ($\mathrm{\mu m}^3$) & (GHz) &  & & & (kHz) & (kHz) & ($\mathrm{\mu m}$)\\
		\colrule
		SR & $770\times 410$ & \num{1.88e7} & $3.7598$ & $12930$ & $31434$ & $21968$ & $438\pm32$ & $511.2$  & $2.53\pm0.82$\\
		MR & $760\times1000$ & \num{1.52e7} & $4.3305$ & $1688$  & $2155$  & $7791$  & $461\pm32$ & $627.6$  & $2.23\pm0.19$\\
		CR & $580\times 800$ & \num{1.41e7} & $4.5398$ & $31262$ & $89694$ & $47987$ & $384\pm8.4$ & $658.0$ & $3.79\pm0.09$\\
	\end{tabular}
	\end{ruledtabular}
    \caption{\label{tab:resonators}Parameters for the three resonator
    geometries, extracted at $T = \SI{1.5}{K}$. The resonance frequency as well
as the quality factors are extracted with a Si:P crystal mounted on the resonator
and at a finite magnetic field. The applied sample power was $P_\mathrm{S} =
\SI{-110}{dBm}$ for the CR and $P_\mathrm{S} = \SI{-115}{dBm}$ for the SR/MR.}
\end{table*}

\subsection{Comparison of the resonator parameters: Experiments vs.\ FEM}
To verify our simulation approach, we fabricated two sample chips with several
capacitively-shunted resonators (CR). We measure the complex transmission
$S_{21}$ with no external applied magnetic field and extract the resonance
frequency as well as the external coupling rate. The measurements were performed
without a sample, therefore we have excluded the additional silicon body on top
of the resonator in the simulations for this section.

\begin{figure}
	\center
	\includegraphics[width=\linewidth]{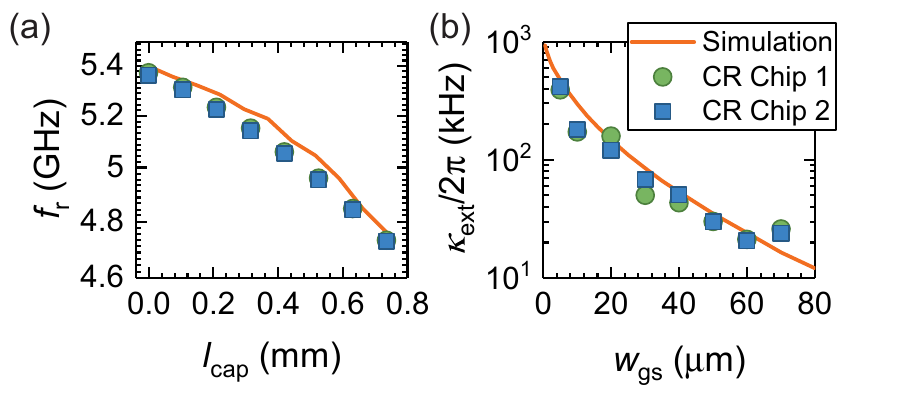}
    \caption{\label{SimulationComparison}Comparison of two chips containing
        several capacitively-shunted resonators (CR) with finite element
        simulations. Filled squares and circles represent experimental data. The
        solid orange lines are obtained from FEM.
    (a)~Tuning of the resonance frequency by adjusting the length
    $l_\mathrm{cap}$ of one of the fingers of the capacitor. The resonance
    frequency drops with increasing length $l_\mathrm{cap}$. 
    (b)~Tuning of external coupling $\kext$ between resonator and feedline by
changing the width $w_\mathrm{gs}$ of the small metal strip, separating the
resonator from the signal line.}
\end{figure}

In Fig.~\ref{SimulationComparison}\,(a) we compare the measured and simulated
resonance frequency as a function of the length of the capacitor finger
$l_\mathrm{cap}$. Increasing $l_\mathrm{cap}$ results in a higher total
capacitance and therefore a decrease in the resonance frequency. The simulations
(orange line) reproduce the measurement results quantitatively within
\SI{1.6}{\%} of the resonance frequency. Nevertheless, the frequency is slightly
overestimated, which we attribute to modeling the superconductor as a perfect
electric conductor neglecting the effects of superconducting properties. 
In panel~(b), the external coupling rate $\kext$ is plotted as a function of the
width of the ground line $w_\mathrm{gs}$, separating the resonator window from
the CPW (see Fig.~\ref{SimulationSetup}\,(b)). As expected, reducing
$w_\mathrm{gs}$ increases the coupling. The coupling rate roughly shows an
exponential behaviour with the separation between the resonator and the
feedline. This is due to the screening of the microwave radiation of the
feedline by a metallized strip with a width $w_\mathrm{gs}$ between the two
circuit elements. Again, we find excellent qualitative agreement between the
experimental data and the finite element modeling.

\subsection{Continuous-wave electron spin resonance}
We perform continuous-wave ESR measurements by placing a phos\-phorus-doped
$^\mathrm{nat}$Si sample with a donor density of $\rho = 2\times
10^{17}\,\mathrm{cm^{-3}}$ in flip-chip geometry on the sample chip. For the
following measurements, both attenuators in the cold part of the microwave
circuitry are in place.

\begin{figure}
	\center
	\includegraphics[width=\linewidth]{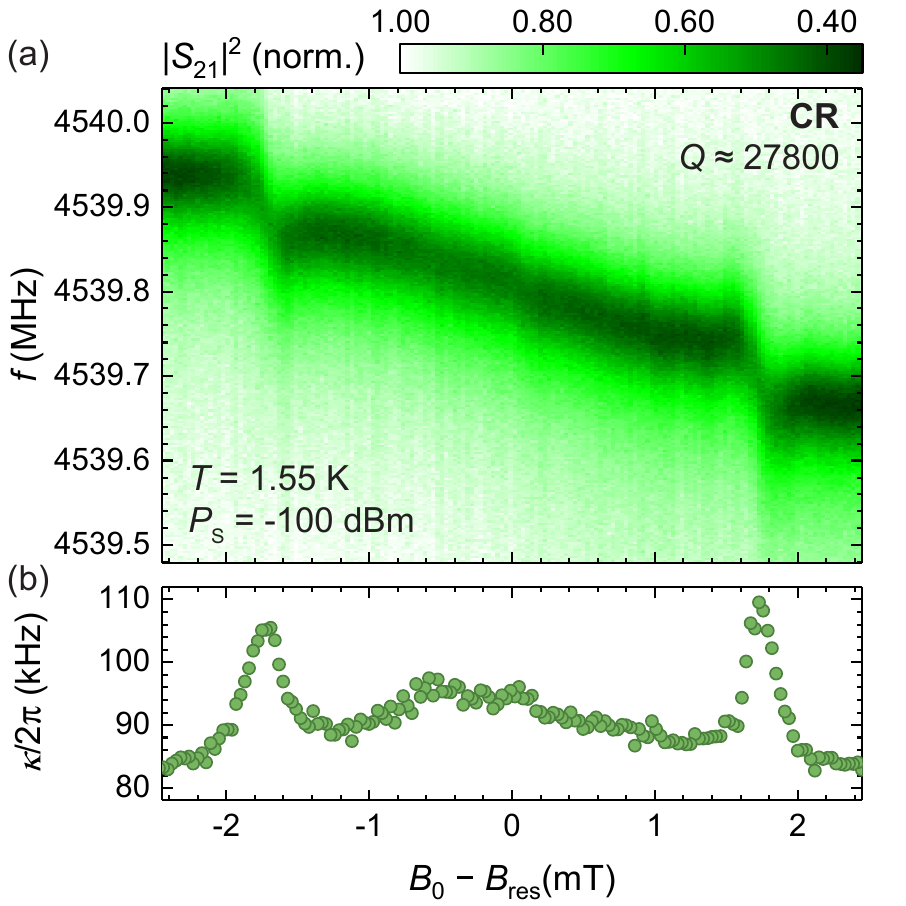}
    \caption{\label{CWMeasurement}(a)~Transmission $\left|S_{21}\right|^2$ as a
        function of frequency and applied magnetic field. The
        frequency-dependent absorption dip corresponds to the resonator, while
        the two distinct features indicate the phosphorus hyperfine transitions.
    (b)~Extracted linewidth $\kappa/2\pi$ (HWHM) as a function of the magnetic
    field. The two peaks correspond to the hyperfine transitions. The features
    at intermediate fields are compatible with Pb$_0$/Pb$_1$ dangling bond
    defects and P$_2$ dimers. (See text for details)
	}
\end{figure}

In Fig.~\ref{CWMeasurement}\,(a) we show the normalized transmission
$|S_{21}|^2$ as a function of probe frequency and the static applied magnetic
field $B_0$ relative to the center resonance field $B_\mathrm{res}$.
The transmission is reduced to about 0.35, when the excitation frequency is in
resonance with the microwave transmission. We further observe a shift of the
resonance frequency over the displayed magnetic field range which we attribute
to the magnetic field dependent kinetic inductance of the superconductor. The
applied magnetic field leads to an increase in kinetic inductance and hereby
also the total inductance of the resonator, changing $f_r$. Further, we observe
two distinct features at $\pm\SI{1.7}{mT}$, which are identified as the two
hyperfine transitions of the phosphorus donors in silicon. 

For a more detailed analysis we determine the resonator linewidth $\kappa$ as a
function of the applied magnetic field. For this, we extract $\kappa$ from the
transmission spectra for each magnetic field step. This corresponds to a
continuous-wave ESR measurement, where the quality factor (absorption signal) of
the resonator is measured\cite{Poole1996}. We plot $\kappa$ as a function of the
applied static field in Fig.~\ref{CWMeasurement}\,(b). In this representation,
the two peaks correspond to the two hyperfine transitions analogously. In the
magnetic field range between the two peaks we observe additional broad features
corresponding to two additional spin systems. The resonance fields of those
peaks are compatible with (i) dangling bond defects at the Si/SiO$_2$ interface,
known as Pb$_0$/Pb$_1$ defects\cite{Poindexter1981,Stesmans1998} ($\approx
\SI{-0.5}{mT}$), and (ii) exchange-coupled donor pairs forming P$_2$
dimers\cite{Feher1955,Jerome1964} ($\approx\SI{0}{mT}$).

Note that we did not perform a field calibration to absolute values.
The static magnetic field in our experiments is generated by a large
superconducting solenoid, which exhibits a significant amount of trapped flux.
This leads to field offsets in the order of $\approx\SI{10}{mT}$. However, in
our work the absolute magnetic field applied to the Si:P crystal is only of
subordinate interest and we therefore plot the magnetic field relative to the
expected center resonance field.
 
For the applied power of $P_\mathrm{S}=\SI{-100}{dBm}$, corresponding to an
average photon number of $n\approx\num{7.8e4}$ (c.f.\
Eq.~\eqref{eq:photon_number}), we observe the onset of saturation effects
(see Section~\ref{sec:power_saturation}). In order to calculate the collective
coupling between the spin ensemble and the microwave resonator in the next
section, we choose a dataset, where the microwave power was decreased to
\SI{-110}{dBm} (\num{2.5e3} photons on average). We also point out the
importance of the additional attenuators in the setup to suppress thermal
microwave noise photons generated at room temperature. Without the attenuation,
we observed the onset of saturation effects already at microwave powers as low
as \SI{-120}{dBm}.

\subsection{Analysis of the collective coupling}
\label{sec:coupling_analysis}
\begin{figure}
	\center
	\includegraphics[width=\linewidth]{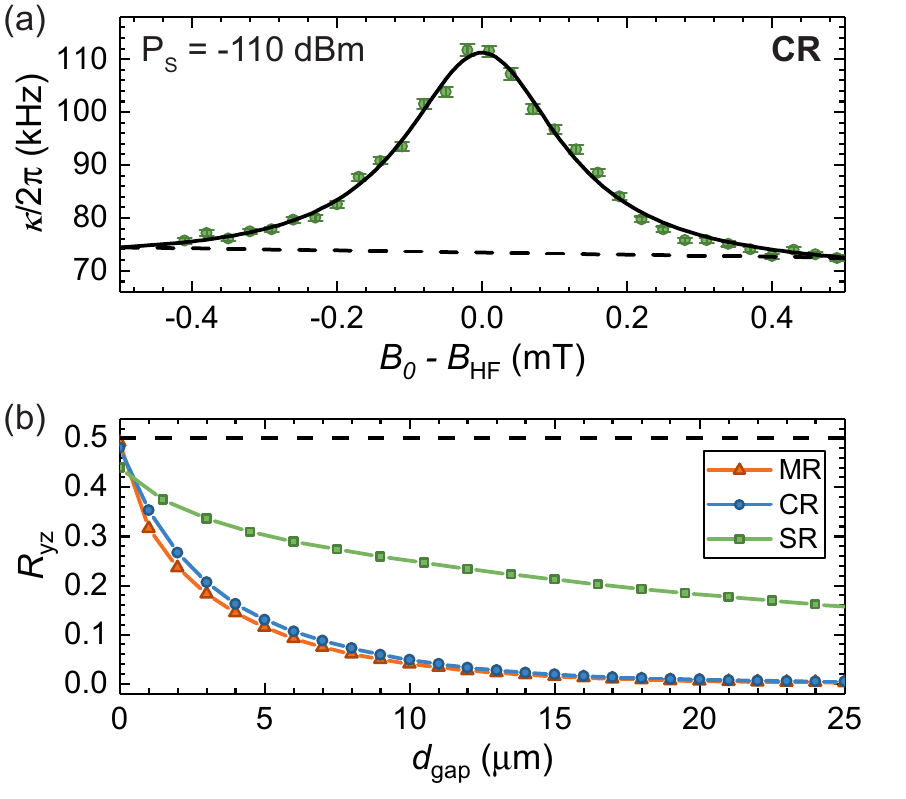}
	\caption{\label{CouplingAnalysis}
        (a) Measured linewidth $\kappa/2\pi$ of the spiral resonator for the
        high-field hyperfine transitions of phosphorus donors in
        $^\mathrm{nat}$Si. The solid line is a fit to
        Eq.~\eqref{eq:herskind} to extract $\geff$, $\kappa_c$ and
        $\gammas$ (see text).
        (b) Dependence of the filling factor $\Ryz$ on a finite gap
    $d_\mathrm{gap}$ between the resonator and the Si:P crystal. A reduction of
$\Ryz$ from the ideal value (dashed line) is due to magnetic field components
parallel to the static magnetic field. The CR and MR show a more significant
decrease compared to the SR due to the short decay length of the dynamic
magnetic field.
}
\end{figure}

In the following we analyze the collective coupling between the microwave
resonator and the high-field hyperfine transition of the phosphorus donors. In
Fig.~\ref{CouplingAnalysis}\,(a) we show the extracted linewidth plotted against
the applied static field, relative to the resonance field of the
high-field hyperfine transition. The microwave power applied to the sample is
$P_\mathrm{S} = -110\,\mathrm{dBm}$ to avoid saturation of the ESR transition. 

The linewidth $\kappa$ in the weak coupling regime can then be described
by\cite{Herskind2009}
\begin{equation}
	\kappa = \kappa_c + \frac{\geff^2\gammas}{\Delta^2 + \gammas^2},
	\label{eq:herskind}
\end{equation}
where $\kappa_c$ is the off-resonant linewidth of the resonator, $\gammas$ is
the spin linewidth (full-width at half maximum) and $\geff$ is the collective
coupling rate. The detuning~$\Delta$ is defined as $\Delta = g_s\muB(B_0 -
B_\mathrm{HF})/h$. We fit Eq.~\eqref{eq:herskind} in combination with a linear
background (dashed line) to the data presented in
Fig.~\ref{CouplingAnalysis}\,(a) (solid lines). We extract a collective coupling
rate $\geff/2\pi = \SI{438\pm32}{kHz}$. The resonator linewidth at $B_0 -
B_\mathrm{HF} = \SI{0.49}{mT}$ is $\kappa/2\pi = \SI{72.4\pm0.5}{kHz}$. The
inhomogeneously
broadened spin linewidth (half-width at half maximum) is $\gammas/2\pi =
\SI{3.67\pm0.13}{MHz}$, corresponding to a linewidth $\delta B =
\SI{131.1\pm4.6}{\mu T}$. This is in agreement with literature values for
$^\mathrm{nat}$Si with a natural abundance of \SI{4.7}{\%} $^{29}$Si
nuclei\cite{Abe2010,Note1}. 
Note that the lineshape of the ESR transition depends on the residual $^{29}$Si
concentration in the sample\cite{Abe2010}. For small $^{29}$Si concentrations,
the lineshape is given by a Lorentzian. However, the transition from Lorentzian
to Gaussian lineshape happens at a $^{29}$Si concentration of
$\approx\SI{5}{\%}$, which is the case for our sample.  For our analysis, we
therefore performed fits with Gaussian and Lorentzian lineshapes. We only
observe a good agreement when fitting a Lorentzian peak, confirming that using
Eq.~\eqref{eq:herskind} is valid.

\footnotetext[1]{We estimate the contribution to the inhomogeneous broadening
due to an inhomogeneous $B_0$ field to be less than \SI{17}{\mu T}, based on the
specified field homogeneity of the solenoid. Due to the way the Si:P crystal is
mounted on the resonator, the influence of strain on the inhomogeneous
broadening is negligible.}

For a first theoretical estimate of $\geff$, we assume an effective spin density
$\rho_\mathrm{eff}=0.5\rho$, as only half of the spins contribute to each
hyperfine transition\cite{Note2}. The thermal spin polarization at $T = 1.55\,\mathrm{K}$
and with the magnetic field on resonance is $1.9\,\%$\cite{Zollitsch2015}. With
these values we obtain for the CR a collective coupling rate
$g_\mathrm{eff,inhom}/2\pi = 645.1\,\mathrm{kHz}$, over-estimating the measured
value by more than $40\,\%$. This deviation can be explained by a finite gap
between the resonator and the Si:P crystal, reducing the effective filling
factor. 

\footnotetext[2]{In our calculation we use the nominal donor concentration
    $\rho$ as the density, assuming that each donor contributes equally to the
    collective coupling. However, for donor concentrations
    $N_D\gtrapprox\SI{5e16}{cm^{-3}}$, phosphorus dimers and trimers are formed,
    which also contribute the the broad background signal and the P$_2$ dimer
    transition. We therefore calculate the theoretically upper bound of the
collective coupling.}

To analyze the dependence of the effective coupling rate on the finite gap size
$d_\mathrm{gap}$, we calculate the filling factor $\Ryz$ for the different
designs by taking only $B_1^{yz}$ for $z\ge d_\mathrm{gap}$ in
Eq.~\eqref{eq:filling_factor} into account. We plot $\Ryz$ as a function of
$d_\mathrm{gap}$ in Fig.~\ref{CouplingAnalysis}\,(b). We observe a qualitative
difference between the CR/MR and the SR. Due to the short decay length of the
dynamic magnetic field for the CR and MR, a finite gap shows a significant
effect on the coupling strength for these two designs. In contrast, the larger
mode volume of the SR leads to a more favorable dependence. Due to the different
ratio of components of the dynamic magnetic field perpendicular to the static
magnetic field, the maximum value for $d_\mathrm{gap}$ differs for the three
designs. We find a maximum value of $R_{yz,\mathrm{SR}} = 0.440$ for the spiral
resonator, $R_{yz,\mathrm{CR}} = 0.481$ for the CR and $R_{yz,\mathrm{MR}} =
0.491$ for the MR.

Using the data presented in Fig.~\ref{CouplingAnalysis}\,(b), we can estimate
the nominal gap between the Si:P crystal and the resonator plane. The measured
collective coupling rate of \SI{384.8}{kHz} corresponds to a gap of
$d_\mathrm{gap,CR} = \SI{3.80\pm0.09}{\mu m}$. We performed the same
analysis of the collective coupling for the SR and MR and present the extracted
parameters in Table~\ref{tab:resonators}. 

From our simulations, we are able to estimate the number of spins addressed in
the measurement using the three-dimensional field distribution and the
collective coupling rate from experiment and obtain $N = \num{4.16e9}$.
Comparing this number to the number of excitations in the resonator using
Eq.~\ref{eq:photon_number}, $n_\mathrm{ph} = \num{7800}$, confirms that we are
in the low-excitation regime.
Since $N\gg n_\mathrm{ph}$, our modelling of the collective coupling is
valid, even though the resonator is not in the ground state\cite{Chiorescu2010}.
It is therefore also applicable at Millikelvin temperatures in the context of
ultra-sensitive solid-state ESR.

\subsection{Power saturation}
\label{sec:power_saturation}
Our theoretical model also allows us to calculate the collective coupling
strength as a function of the applied microwave power and to take the effects of
power saturation into account. In electron spin resonance, power saturation is a
well-known phenomenon\cite{Castner1959,Cullis1976} which leads to a modulation
of the ESR signal as a function of the applied microwave power. As the power
level increases, the spin system is driven into saturation. This has also an
effect on the collective coupling rate $\geff$, as it leads to an effective
decoupling of the spin system from the microwave resonator\cite{Angerer2017}.

\begin{figure}
	\center
	\includegraphics[width=\linewidth]{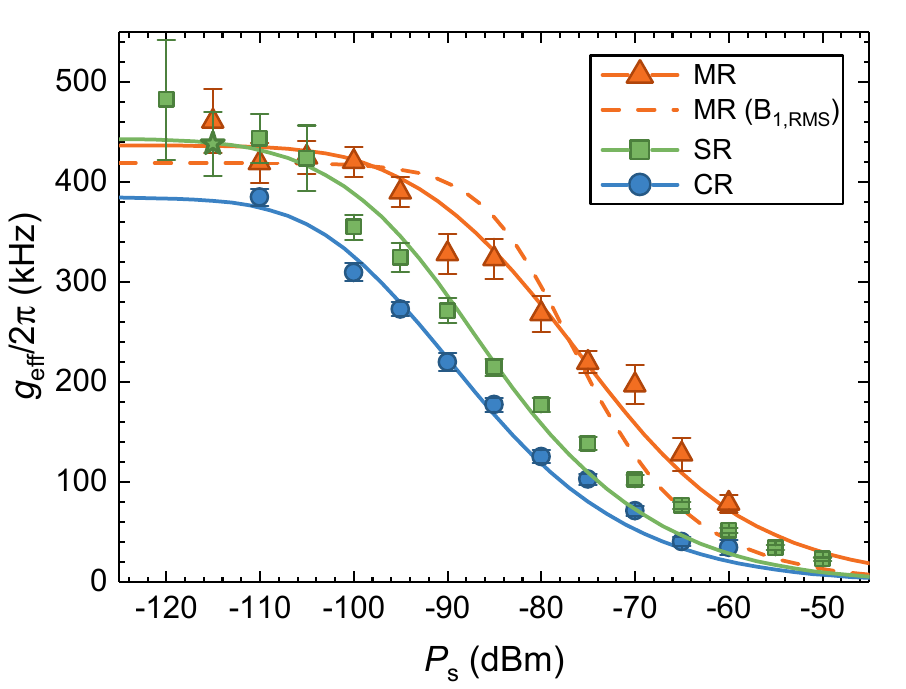}
    \caption{\label{geff_vs_power}Effective coupling rate $\geff$ as a function
    of the applied power to the sample $P_\mathrm{S}$. The effective coupling
rate $\geff$ decreases with increasing microwave power, indicating a
power-dependent saturation effect. The solid lines are fits to
Eq.~\eqref{eq:g_eff_inhom_sat}, taking the simulated magnetic field distribution
into account. The dashed line is a calculation using the mean field amplitude
$B_{1,\mathrm{rms}}$ of the CR.}
\end{figure}

In Fig.~\ref{geff_vs_power} we plot the collective coupling rate as  a function
of the applied power for the three resonator geometries. The collective coupling
rates were obtained using the procedure described in the previous section. The
SR and MR achieve the highest coupling rates of about \SI{450}{kHz}. All three
curves show a decrease of the collective coupling with increasing  power,
suggesting a power-dependent saturation effect.

In order to model the power dependence, we take the saturation of the ESR signal
into account.
In a continuous-wave ESR experiment, the signal increases with the microwave
driving field $B_1$ and the signal intensity is therefore given
by\cite{Fajer1992}
\begin{equation}
	S = \frac{S_0B_1}{(1 + s\gamma^2 B_1^2 T_1 T_2)^{1/2}}.
	\label{eq:saturation_factor}
\end{equation}
Here, $\gamma$ is the gyromagnetic ratio and $T_1$ and $T_2$ are the spin life
time and spin coherence time, respectively. The exponent of $1/2$ is valid for
an inhomogeneously broadened spin ensemble, which is the case for
$^\mathrm{nat}$Si with abundant $^{29}$Si nuclei\cite{Abe2010}. $S_0$ is a
factor independent of $B_1$ and the factor $S_\mathrm{sat} = 1/(1 + \gamma^2
B_1^2 T_1 T_2)^{1/2}$ models the power saturation. For $\gamma^2 B_1^2 \ll
1/(T_1 T_2)$, the denominator in Eq.~\eqref{eq:saturation_factor} is equal to one
and the spin ensemble is in the unsaturated regime. The correction factor $s$
accounts for Lorentzian or Gaussian lineshapes and is determined via a
least-squares fitting procedure. For the $B_1$-field in this equation we use the
exported field distribution from our simulation data. We rescale the magnetic
field in Eq.~\eqref{eq:saturation_factor} according to $B_1 \propto
\sqrt{P_\mathrm{S}}$ to obtain a power-dependent expression. Using this result,
we can modify Eq.~\eqref{eq:filling_factor} to take power saturation into
account and obtain
\begin{equation}
	R_\mathrm{yz,sat} = \frac{\sum_{V} |S_\mathrm{sat}(B^{yz}_{1,\mathrm{sim}}(\vec{r}))\cdot B^{yz}_{1,\mathrm{sim}}(\vec{r})|^2}{\sum_{V_\mathrm{m}} |B^{xyz}_{1,\mathrm{sim}}(\vec{r})|^2}.
	\label{eq:g_eff_inhom_sat}
\end{equation}
The solid lines in Fig.~\ref{geff_vs_power} are fits of
Eq.~\eqref{eq:g_eff_inhom_sat} to the data with the correction factor $s$ being
the only free parameter. The values of $T_1 = \SI{48}{ms}$ and $T_2 =
\SI{400}{ns}$ are extracted from pulsed ESR measurements. We obtain a good
agreement of the data with our model if the full field distribution is taken
into account. Using only the root-mean square amplitude of the field
distribution, we do obtain a significantly worse agreement between theory and
experiment (dashed line).
The good agreement of our model regarding the power saturation
behaviour again confirms our understanding of spatial extent of the microwave
excitation field.

\section{Discussion and conclusion}
In this work, we presented three different resonator designs. All designs offer
distinct advantages, and therefore choosing the appropriate resonator geometry
depends on the intended application. Superconducting planar resonators are used
in two different experimental configurations with a different interface between
the spin ensemble and the microwave resonator. First, one widely used method is
the flip-chip configuration, where the spin sample is placed on top of the
resonator, while the resonator itself is placed onto a separate substrate
(e.g.~Ref.~\citenum{Kubo2010,Amsuss2011,Probst2013,Zollitsch2015} and this
work). This type of sample mounting offers greater experimental
flexibility, as the sample preparation and placement is independent of the
fabrication of the resonator. However, due to the flip-chip geometry, the
presence of a small gap between the microwave resonator and the spin sample is
highly likely, in particular when working with solid state samples. Another
option is to pattern the microwave resonator directly onto a substrate which
already contains the spin ensemble
(e.g.~Ref.~\citenum{Bienfait2016,Eichler2017,Probst2017}). Here, the direct
interface between the spin ensemble and the microwave resonator offers the
largest spin-photon coupling rates but comes with the disadvantage of
strain-induced spin resonance shifts\cite{Pla2018}. The latter are e.g.\ caused
by different lattice constants and thermal expansion coefficients of substrate
and the superconducting thin film material of the resonator.

From our simulation and experimental results, we can draw two conclusions.
First, the simulated field distribution of the SR shows that this design is
favorable for a homogeneous excitation of the spin ensemble in a flip-chip
geometry. The microwave magnetic field decays slowly along the $z$-direction and
therefore offers spin excitation throughout the whole spin sample region, even
for bulk samples. This makes this design more robust against the presence of a
gap between the resonator and the spin sample. For better controlled
resonator-to-spin sample distance and an even more enhanced field homogeneity, a
thin polyimid (Kapton) spacer can be inserted between the microwave resonator
and the spin sample to avoid $B_1$ inhomogeneities in the near-field of the
resonator\cite{Benningshof2013}. From the simulations of the SR we find an
extended region of \SI{20}{\mu m} thickness with an homogeneity better than
\SI{10}{\%}, comparable to commercial microwave resonators.

Second, if high single spin coupling rates are desired, a resonator with a
meander-shaped strip patterned directly on top of a thin layer containing the
spin ensemble offers the best performance. In particular, the CR design with its
large finger capacitor and therefore small inductance offers a high current
density and therefore a large $B_1$ field, increasing the single spin coupling
rate\cite{Eichler2017}. Additionally, the periodic pattern of the meandering
wire allows to engineer microwave antennas that emit microwaves with a specific
wavevector. This is a concept that is widely used in spin wave resonance
spectroscopy, where excitation at a non-zero wave vector is
desirable\cite{Bailleul2003}.

To summarize, we have analyzed three designs for superconducting planar lumped
element resonators. We performed finite element method based simulations to
extract the resonance frequency and the quality factor as well as to calculate
the characteristic magnetic field distribution of each resonator. We obtained a
good agreement between the simulations and the experimental results of
fabricated chips, where the resonators were structured into a $150\,\mathrm{nm}$
Nb film. The spiral resonator exploits the two-dimensional coil-like arrangement
of the resonator wire to obtain an improved magnetic field homogeneity as well
as an increased filling factor when a finite gap between the Si:P crystal and
resonator is present. To demonstrate the feasibility of the resonators for
magnetic resonance experiments, we presented continuous-wave ESR measurements on
phosphorus-doped $^\mathrm{nat}$Si using all three resonator designs. In order
to explain the extracted collective coupling rates, we extended our existing
theoretical model to take the simulated microwave magnetic field distribution
into account and found a good agreement between the data and the model. Finally,
we extracted the collective coupling rate as a function of applied microwave
power and modeled the saturation behavior using our model. Our research
demonstrates the feasibility of finite element method based simulations to
extract the expected collective coupling rate of a spin ensemble coupled to a
microwave resonator and gives insight into the different application cases of
the resonator designs.

\begin{acknowledgments}
We thank stimulating discussions with S.\,T.\,B.~Goennenwein and M.\,S.~Brandt.
We acknowledge financial support from the German Research Foundation via
SPP~1601~(HU~1896/2\mbox{-}1).
\end{acknowledgments}

%

\appendix
\section{Collective coupling in an inhomogeneous magnetic field}
\label{sec:g_eff_derivation}
In the following we derive an expression allowing us to calculate the collective
coupling rate $\geff$ for a spatially inhomogeneous distribution of the dynamic
magnetic field. We first start with the case of a homogeneous field.

The collective coupling strength of a spin ensemble of $N$ spins is given
by\cite{Sandner2012} $\geff = \sqrt{\sum_j^N |g_j|^2}$, where $g_j$ is the
single-spin coupling strength of an individual spin in the ensemble. In the case
of homogeneous coupling, this expression reduces to $\geff = g_0\sqrt{N}$ with
the single-spin coupling rate given by\cite{Wesenberg2009}
\begin{equation}
	g_0 = \frac{g_s\muB}{2\hbar}B_{1,0}.
	\label{eq:g_0_hom}
\end{equation}
Here, the coupling strength is determined by the magnetic component of the
microwave vacuum field, $B_{1,0}$. This field can be estimated by integrating
the energy stored in the magnetic field created by vacuum fluctuations,
according to\cite{Schoelkopf2008}
\begin{equation}
	\frac{\hbar\omega_r}{4} = \frac{1}{2\mu_0}\int_{V_\mathrm{m}} B_1^2\,\mathrm{dV} = \frac{B_{1,0}^2 V_\mathrm{m}}{2\mu_0}.
	\label{eq:b_1_vacuum_hom}
\end{equation}
The additional factor of $2$ in the denominator on the left side of the equation
takes into account that half of the energy is stored in the electric component
of the vacuum field. 
Using Eq.~\eqref{eq:g_0_hom}) and \eqref{eq:b_1_vacuum_hom} the collective
coupling strength for a spatially homogeneous field distribution is given by
\begin{equation}
	g_\mathrm{eff,hom} = \frac{g_s\muB}{2\hbar}\sqrt{\frac{\mu_0\rho_\mathrm{eff}\hbar\omega_r \nu}{2}},
	\label{eq:g_eff_hom_app}
\end{equation}
where we have substituted $N = \rho_\mathrm{eff} V$ with the effective spin
density $\rho_\mathrm{eff}$ and the spin sample volume $V$. The ratio $\nu =
V/V_\mathrm{m}$ is called filling factor and defines the volume of the resonator
field filled with the spin ensemble.

In the case of planar resonators, the assumption of a spatially homogeneous
$B_1$ field is not valid. Here, we have to take into account the spatial
dependence of $B_1(\vec{r})$ in Eq.~\eqref{eq:b_1_vacuum_hom}. Another point to
consider is the orientation of the $B_1$ field with respect to the static
magnetic field $\vec{B_0}$. In our experiments the static field is applied
in-plane along the $x$-axis (c.f.\ Fig.~\ref{SimulationSetup}). Only field
components $B_1^{yz}$ perpendicular to $\vec{B_0}$ can excite spins in the spin
ensemble. From our simulations we export the three-dimensional magnetic field
$B_{1,\mathrm{sim}}(\vec{r})$, discretized in finite elements with volume
$\Delta V = \Delta x\Delta y\Delta z$. Assuming the field is homogeneous over
the spatial extent of a single volume element we find for the collective
coupling strength of a single volume element 
\begin{equation}
	\tilde{g}_\mathrm{eff}(\vec{r}) = \frac{g_s\muB}{2\hbar}(\alpha B^{yz}_{1,\mathrm{sim}}(\vec{r}))\sqrt{\rho_\mathrm{eff}\Delta V}
	\label{eq:g_eff_volelem}
\end{equation}
where $B^{yz}_{1,\mathrm{sim}}(\vec{r})$ is the exported field amplitude of the
volume element and $\rho_\mathrm{eff}\Delta V$ is the number of spins in the
volume element. The calibration factor $\alpha$ rescales the field amplitude
from the simulated excitation level to the level of vacuum fluctuations. It can
be calculated similar to Eq.~\eqref{eq:b_1_vacuum_hom} by
\begin{equation}
	\frac{\hbar\omega_r}{4} = \frac{1}{2\mu_0}\int_{V_\mathrm{m}} (\alpha B^{xyz}_{1,\mathrm{sim}}(\vec{r}))^2\,\mathrm{dV}.
\end{equation}
Note that here we also take the $x$-component of the $B_1$ field into account.
We convert the integration into a summation over all volume elements and find
for the calibration factor
\begin{equation}
	\alpha = \sqrt{\frac{\mu_0\hbar\omega_r}{2\Delta V\sum_{\Delta V} (B^{xyz}_{1,\mathrm{sim}}(\vec{r}))^2}}.
	\label{eq:alpha}
\end{equation}

Combining Eq.~\eqref{eq:g_eff_volelem} and \eqref{eq:alpha} and summing over
all volume elements we finally obtain an expression for the collective coupling
strength of a spin ensemble coupled to an inhomogeneous microwave magnetic
field:
\begin{align}
	\geff \nonumber &= \sqrt{\sum_{\Delta V} |\tilde{g}_\mathrm{eff}|^2} =\\ 
	               &= \frac{g_s\muB}{2\hbar}\sqrt{\frac{\mu_0\hbar\omega_r\rho_\mathrm{eff}}{2}}\sqrt{\frac{\sum_{\Delta V} |B^{yz}_{1,\mathrm{sim}}(\vec{r})|^2}{\sum_{\Delta V} |B^{xyz}_{1,\mathrm{sim}}(\vec{r})|^2}}.
\end{align}
Note that in this expression the size of the volume element $\Delta V$
effectively cancels out. The last square-root term plays a similar role to the
filling factor $\nu$ introduced in Eq.~\eqref{eq:g_eff_hom_app}.

\end{document}